\documentclass[aps,prd,reprint,superscriptaddress]{revtex4-2}

\usepackage{amsmath}
\usepackage{bbold}
\usepackage{amsfonts}
\usepackage{amssymb}
\usepackage{pbsi}
\usepackage[T1]{fontenc}
\usepackage{xcolor}
\usepackage{graphicx}
\usepackage{natbib}
\usepackage{float}
\usepackage{color}
\usepackage{hyperref}
\hypersetup{colorlinks = true,
	linkcolor = blue,
	anchorcolor = blue,
	citecolor = blue,
	filecolor = blue,
	urlcolor = blue}

\begin{document}
\title{Reconnection of Gravitational Fields}

\author{Luca Comisso}
\email{luca.comisso@columbia.edu}
\affiliation{Department of Astronomy, Columbia University, New York, NY 10027, USA} 
\affiliation{Department of Physics, Columbia University, New York, NY 10027, USA}
\affiliation{Columbia Astrophysics Laboratory, Columbia University, New York, NY 10027, USA}

\author{Felipe A. Asenjo}
\email{felipe.asenjo@uai.cl}
\affiliation{Facultad de Ingenier\'ia y Ciencias, Universidad Adolfo Ib\'a\~nez, Santiago 7491169, Chile}

\begin{abstract}

$\,$

In a tetrad formulation of general relativity, an ideal Ohm-type condition for the gravitational field implies the conservation of gravitational fluxes and the preservation of the connectivity of gravitational field sheets. 
We investigate the consequences of departures from ideal gravitational evolution by developing a theory of gravitational reconnection, defined as a change in the connectivity of gravitational field sheets arising from the breakdown of the ideal frozen-in evolution of the gravitational field.
We derive quantitative measures of the gravitational reconnection rate, show that gravitational reconnection alters the energy-momentum exchange between the gravitational and matter sectors, and determine how it modifies the evolution of gravitational helicity.
Our results provide a quantitative framework for describing changes in gravitational connectivity and their consequences in dynamical spacetimes.

$\,$

$\,$

\end{abstract}

\maketitle

\section{Introduction}

General relativity provides the theoretical framework for describing dynamical spacetimes, including those associated with black-hole and neutron-star mergers, core-collapse supernovae, and the formation of large-scale cosmological structure.
One of the main objectives is to develop a predictive and physically transparent description of these spacetimes. 
Much of the progress toward this goal has relied on numerical relativity simulations \cite{Lehner01,Alcubierre08,BS10}, which have enabled detailed studies of compact-object mergers \cite{Pretorius2005,Campanelli2006,Baker2006} and the interpretation of gravitational-wave observations \cite{Abbott2016,Abbott2017}. Numerical relativity has thus become the principal tool for investigating nonlinear spacetime dynamics. At the same time, the complexity of these spacetimes motivates the search for general principles that go beyond individual solutions and characterize the geometric structure of the gravitational field.

This search for general principles lies at the heart of geometrodynamics \cite{Wheeler63,Wheeler64}, which seeks to understand gravitation through the dynamical geometry of spacetime \cite{MTW,Thorne12}. 
An important development within this program has been the reformulation of the Einstein equations in terms of quantities analogous to electric and magnetic fields \citep[e.g.][]{MaarBass98}, whereby spacetime curvature is organized into field-like geometric structures with an electrodynamical interpretation. 
Field-based descriptions of gravitation have found applications in the visualization and analysis of tidal gravitational accelerations and differential frame dragging through tendex and vortex lines \cite{Owen11,Nichols11,Nichols12,TB17}, studies of relativistic tidal dynamics around compact objects \cite{Poisson05,Poisson10,Camilloni23}, as well as in Maxwell-type interpretations of binary-black-hole mergers and gravitational-wave turbulence \cite{boye,Krynicki}. 
These developments underscore the value of electrodynamical formulations of general relativity for analyzing nonlinear spacetime dynamics.

Of particular importance for the present work, an electrodynamical formulation of general relativity, together with an ideal Ohm-type condition for the gravitational field, reveals a set of topological constraints governing spacetime dynamics \cite{AWC}. These constraints parallel those governing force-free electrodynamics \cite{Carter79,Uchida97a,Uchida97b,Gralla14} and dissipationless magnetohydrodynamics \cite{AC15,pegoraro16,magnetoconnection1,magnetoconnection2}.
Under an ideal evolution of the gravitational field, the connectivity of gravitational field sheets is preserved, and gravitational flux and gravitational helicity satisfy corresponding conservation laws.
These frozen-in properties and conservation laws constrain the admissible spacetime dynamics and provide the basis for defining the motion, connectivity, and topology of gravitational field sheets \cite{AWC}.

Understanding spacetime dynamics requires understanding not only the ideal evolution governed by these topological constraints, but also how these constraints break down. When the ideal frozen-in evolution is violated, the connectivity of gravitational field sheets can change through their reconfiguration. By analogy with magnetic reconnection in plasmas, we refer to this process as gravitational reconnection.
In the present work, we formulate a theory of gravitational reconnection. We identify the conditions under which gravitational connectivity changes, introduce quantitative measures of the gravitational reconnection rate, and examine its implications for energy-momentum exchange between the gravitational and matter sectors and for gravitational helicity. These results provide a framework for describing gravitational reconnection and its consequences in dynamical spacetimes.

Throughout this paper, we adopt units with $G=c=1$ and metric signature $(-,+,+,+)$.

\section{Gravitational Field Equations}

Einstein's field equations,
\begin{equation} \label{eq:EFE}
G_{\mu\nu}=8\pi T_{\mu\nu} \, ,
\end{equation}
can be recast in a form analogous to nonlinear electrodynamics in continuous media. This becomes manifest in a tetrad formulation of general relativity, characterized by a local tetrad satisfying
\begin{equation}
\mathcal{A}^\mu{}_{\hat{\alpha}} \, \mathcal{A}^\nu{}_{\hat{\beta}} \, g_{\mu\nu} = \eta_{\hat{\alpha}\hat{\beta}}  \, ,
\end{equation}
with $g_{\mu\nu}$ the spacetime metric and $\eta_{\hat\alpha\hat\beta}$ the Minkowski metric. Unhatted Greek indices $\mu,\nu,\lambda,\ldots$ label spacetime components, while hatted Greek indices $\hat\alpha,\hat\beta,\hat\gamma,\ldots$ label tetrad-frame components. 
Treating the tetrad as a gravitational vector potential \cite{Olivares22,Peshkov25}, the corresponding gravitational field-strength tensor is
\begin{equation}
\mathcal{F}^{\hat{\alpha}}{}_{\mu\nu} = \partial_\mu \mathcal{A}^{\hat{\alpha}}{}_\nu - \partial_\nu \mathcal{A}^{\hat{\alpha}}{}_\mu \, .
\label{eq:field_strength_tensor}
\end{equation} 
For a torsion-free, metric-compatible connection, the spin connection takes the form 
\begin{align}
\omega_{\hat{\alpha}\hat{\beta}\hat{\gamma}}
= \frac{1}{2} \left( \mathcal{F}_{\hat{\beta}\hat{\alpha}  \hat{\gamma}} + \mathcal{F}_{\hat{\gamma}\hat{\alpha}  \hat{\beta}} - \mathcal{F}_{\hat{\alpha}\hat{\beta}\hat{\gamma}}\right) \, ,
\end{align}
where
\begin{equation}
\mathcal F^{\hat\alpha}{}_{\hat\beta\hat\gamma}
=
\mathcal A^\mu{}_{\hat\beta}
\mathcal A^\nu{}_{\hat\gamma}
\mathcal F^{\hat\alpha}{}_{\mu\nu} \, .
\end{equation}

Central to the reformulation of the Einstein equations is the dual Nester-Witten form \cite{Nester81,Witten81}, defined in terms of the spin connection by
\begin{align}
^*{\mathcal{U}}^{\hat{\beta}\hat{\gamma}}{}_{\hat{\alpha}} = \omega^{[\hat{\beta}\hat{\gamma}]}{}_{\hat{\alpha}} + \delta^{\hat{\beta}}{}_{\hat{\alpha}} \omega^{[\hat{\gamma}\hat{\delta}]}{}_{\hat{\delta}} - \delta^{\hat{\gamma}}{}_{\hat{\alpha}} \omega^{[\hat{\beta}\hat{\delta}]}{}_{\hat{\delta}} \, .
\end{align} 
The dual Nester–Witten form allows the tetrad-projected Einstein tensor to be represented as a Levi-Civita covariant divergence term plus a self-interaction term \cite{Frauendiener90,Szabados92},
\begin{equation}
G^{\mu}{}_{\hat{\alpha}}
= \nabla_{\nu} {^*{\mathcal{U}}_{\hat{\alpha}}}{}^{\nu\mu} - t^{\mu}{}_{\hat{\alpha}} \, ,
\label{eqn:G_NW}
\end{equation} 
where the gravitational Sparling self-current $t^{\mu}{}_{\hat{\alpha}}$ is given by  
\begin{equation}
{t^{\hat\mu}}_{\hat\nu}={\mathcal{F}^{\hat\alpha}}_{\hat\beta\hat\nu}\, {{^*{\cal U}}_{\hat\alpha}}^{\hat\beta\hat\mu}- \frac{1}{4} {\delta^{\hat\mu}}_{\hat\nu} {\mathcal{F}^{\hat\alpha}}_{\hat\beta\hat\lambda}\, {{^*{\cal U}}_{\hat\alpha}}^{\hat\beta\hat\lambda} \, .
\label{eqn:selfcurrent}
\end{equation}
Here and throughout, $\nabla_{\nu}$ denotes the Levi-Civita covariant derivative acting only on spacetime indices.

With the Sparling--Nester--Witten decomposition, Eqs.~\eqref{eqn:G_NW} and \eqref{eqn:selfcurrent}, Einstein's field equations \eqref{eq:EFE} take the Sparling form \cite{Sparling82,Frauendiener90,Szabados92}
\begin{equation} 
\nabla_\mu\,  {^*{\cal U}_{\hat\alpha}}^{\mu\nu} = {t_{\hat\alpha}}^\nu + 8\pi \, {T_{\hat\alpha}}^\nu \, .
\label{einsteinEqs2}
\end{equation}
The Sparling equation is complemented by the homogeneous identity following from the definition \eqref{eq:field_strength_tensor},
\begin{equation}
\nabla_\mu\,  {{^*\mathcal{F}}_{\hat\alpha}}^{\mu\nu}=0\, ,
\label{homogeneous}
\end{equation}
where ${^*\mathcal{F}}^{\hat{\alpha}}{}_{\mu\nu} = \frac{1}{2}\,\epsilon_{\mu\nu\rho\sigma}\, \mathcal{F}^{\hat{\alpha}\rho\sigma}$ is the dual gravitational field-strength tensor. 
Equation \eqref{einsteinEqs2} is equivalent to the Einstein field equations and has a structure closely analogous to the inhomogeneous Maxwell equations. Since ${^*{\cal U}_{\hat\alpha}}^{\mu\nu}$ is antisymmetric in its spacetime indices, $\nabla_\nu\nabla_\mu {^*{\cal U}_{\hat\alpha}}^{\mu\nu}=0$ identically, implying the conservation law 
\begin{equation}
\nabla_\nu ({t_{\hat\alpha}}^\nu + 8\pi \, {T_{\hat\alpha}}^\nu)=0 
\label{cons_current}
\end{equation}
for the combined gravitational plus matter energy-momentum current.

\section{Ideal Evolution of the Gravitational Field}

The electrodynamical formulation of the Einstein equations, together with an ideal Ohm-type condition for the gravitational field, implies the conservation of gravitational flux associated with the field-strength two-form $\mathcal F^{\hat\alpha} = \frac{1}{2} {\mathcal F^{\hat\alpha}}_{\mu\nu} d x^{\mu} \wedge d x^{\nu}$, as well as the preservation of the connectivity of the associated gravitational field sheets.
As the hatted Lorentz index $\hat\alpha$ labels the four field-strength two-forms associated with the tetrad, the following discussion applies independently to each member of this collection.

Let $u^\mu$ be a timelike transport four-velocity.
Assume that the gravitational field strength satisfies an ideal Ohm-type condition \cite{AWC}
\begin{equation}
u^\lambda {\mathcal{F}^{\hat\alpha}}_{\lambda\nu}=0\, .
   \label{ideal_ohm}
\end{equation} 
Equation~\eqref{ideal_ohm} is the direct analogue of the ideal Ohm law of relativistic magnetohydrodynamics, with the electromagnetic field-strength tensor replaced by the gravitational field-strength tensor. 
For a tetrad adapted to the transport congruence, whose timelike leg is aligned with $u^\mu$, so that $\mathcal A^{\hat\alpha}{}_\mu u^\mu=\delta^{\hat\alpha}{}_{\hat0}$,
Eq.~\eqref{ideal_ohm} can also be interpreted as a condition on the propagation of the adapted tetrad. A similar interpretation based on tetrads adapted to freely falling observer congruences is adopted in \cite{Krynicki}.

The gravitational field strength defined by Eq.~\eqref{eq:field_strength_tensor} satisfies identically $\nabla_{[\lambda} {\mathcal F^{\hat\alpha}}_{\mu\nu]} = 0$. Contracting it with the transport four-velocity $u^\lambda$ yields
\begin{equation}
u^\lambda
\nabla_\lambda
{\mathcal F^{\hat\alpha}}_{\mu\nu}
= - u^\lambda
\nabla_\mu
{\mathcal F^{\hat\alpha}}_{\nu\lambda}
- u^\lambda
\nabla_\nu
{\mathcal F^{\hat\alpha}}_{\lambda\mu} \, .
\label{eq:transport_identity}
\end{equation}
Using Eq.~\eqref{eq:transport_identity} together with
Eq.~\eqref{ideal_ohm} yields
\begin{equation}
(\mathcal L_u \mathcal F^{\hat\alpha})_{\mu\nu}=0 \, ,
\label{lieproof}
\end{equation}
where $\mathcal L_u$ denotes the Lie derivative along $u^\mu$, 
\begin{equation}
(\mathcal L_u \mathcal F^{\hat\alpha})_{\mu\nu}
=
u^\lambda \nabla_\lambda {\mathcal F^{\hat\alpha}}_{\mu\nu}
+ {\mathcal F^{\hat\alpha}}_{\lambda\nu} \nabla_\mu u^\lambda
+ {\mathcal F^{\hat\alpha}}_{\mu\lambda} \nabla_\nu u^\lambda \, .
\end{equation}
The gravitational field-strength two-form $\mathcal F^{\hat\alpha}$ is thus Lie transported by the flow generated by $u^\mu$.

The Lie-transport equation \eqref{lieproof} implies the conservation of the gravitational flux through two-surfaces transported by the flow generated by $u^\mu$. Let $\mathcal S(\tau)$ denote a one-parameter family of such two-surfaces, and define the gravitational flux as
\begin{equation}
\Phi^{\hat\alpha}(\tau)
=\frac{1}{2}\int_{\mathcal S(\tau)}
{\mathcal F^{\hat\alpha}}_{\mu\nu}
\,d\Sigma^{\mu\nu} \, , 
\label{flux_tensor_def}
\end{equation}
where $d\Sigma^{\mu\nu}$ is the oriented antisymmetric area element on $\mathcal S(\tau)$. 
By the transport theorem for differential forms,
\begin{equation}
\frac{d\Phi^{\hat\alpha}}{d\tau} =
\frac{1}{2}\int_{\mathcal S(\tau)}
(\mathcal L_u \mathcal F^{\hat\alpha})_{\mu\nu}
\,d\Sigma^{\mu\nu} \, . \label{eq:flux_transport}
\end{equation}
It follows that 
\begin{equation}
\frac{d\Phi^{\hat\alpha}}{d\tau} = 0 \, .
\end{equation}
Thus, the flow generated by $u^\mu$ preserves gravitational flux.

The Lie transport of the gravitational field-strength two-form implies both the conservation of gravitational flux through every transported two-surface and the preservation of the connectivity of the associated gravitational field sheets. 
More precisely, denoting by $d\ell^\mu=x'^\mu-x^\mu$ the infinitesimal separation vector joining neighboring integral curves of $u^\mu$, such that 
\begin{equation}
u^\lambda \nabla_\lambda d\ell^\mu=d\ell^\lambda \nabla_\lambda u^\mu \, ,
\end{equation} 
and using Eq.~\eqref{lieproof}, we obtain the gravitational field-connection equation \cite{AWC}
\begin{equation}
u^\lambda \nabla_\lambda
\left(d\ell^\mu{\mathcal F^{\hat\alpha}}_{\mu\nu}\right)
=
-\left(\nabla_\nu u^\lambda\right)\left(d\ell^\mu{\mathcal F^{\hat\alpha}}_{\mu\lambda}\right) \, ,
\label{connThe4}
\end{equation} 
which governs the preservation of the connectivity of the gravitational field sheets. Equation \eqref{connThe4} implies that if
\begin{equation}
d\ell^\mu {\mathcal F^{\hat\alpha}}_{\mu\nu} = 0 
\label{connThe5}
\end{equation}
is satisfied on an initial two-surface, then it remains satisfied under transport by the flow generated by $u^\mu$, provided that the transport field remains regular.
Consequently, the directions annihilated by $\mathcal F^{\hat\alpha}{}_{\mu\nu}$ define a family of integral two-surfaces, which we refer to as gravitational field sheets, whose connectivity remains invariant under transport by the flow generated by $u^\mu$.

\section{Gravitational Reconnection}\label{sec:grav_reconn}

We define \emph{gravitational reconnection} as a change in the connectivity of gravitational field sheets. Since the ideal gravitational Ohm-type condition~\eqref{ideal_ohm}, together with a regular transport field $u^\mu$, preserves the connectivity of gravitational field sheets, any gravitational reconnection process necessarily requires either a breakdown of Eq.~\eqref{ideal_ohm} or the occurrence of a singularity in the transport field.

To describe departures from ideal transport, we relax Eq.~\eqref{ideal_ohm} and consider a generalized (non-ideal) gravitational Ohm-type condition 
\begin{equation}
u^\lambda \mathcal{F}^{\hat\alpha}{}_{\lambda\nu} = \mathcal{R}^{\hat\alpha}{}_{\nu} \, ,
\label{general_ohm_law}
\end{equation}
where $\mathcal{R}^{\hat\alpha}{}_{\nu}$ encodes departures from ideal transport. Owing to the antisymmetry of $\mathcal{F}^{\hat\alpha}{}_{\mu\nu}$, it necessarily satisfies $u^\nu \mathcal{R}^{\hat\alpha}{}_{\nu}=0$.
Even an infinitesimally small non-ideal term can, in principle, permit gravitational reconnection.

A nonvanishing $\mathcal{R}^{\hat\alpha}{}_{\nu}$ may have different origins. 
It may arise when no transport congruence exists for which Eq.~\eqref{ideal_ohm} is satisfied. 
Singularities, including those associated
with black-hole spacetimes, may constitute one possible source of non-ideal behavior. Other structures, such as topological defects, may also be relevant if they induce a departure from the ideal gravitational Ohm-type condition.
Another possibility is that $\mathcal R^{\hat\alpha}{}_{\nu}$ represents an effective coarse-grained description in which unresolved degrees of freedom are encoded in a non-ideal constitutive relation. These unresolved degrees of freedom may belong to either the matter or gravitational sector. In the gravitational sector, they may include gravitational-wave modes on scales not resolved by the effective description. Quantum effects may also enter such an effective description. For example, quantum-field effects associated with Hawking radiation can influence the spacetime geometry through their backreaction.
Finally, modifications of general relativity, such as $f(R)$ gravity or theories involving Lorentz violation in the gravitational sector, alter the gravitational dynamics relative to general relativity. If such modifications produce a departure from the ideal gravitational Ohm-type condition, their
effect could in principle be represented phenomenologically by $\mathcal{R}^{\hat{\alpha}}{}_{\nu}$.
Here, rather than specifying a particular form of $\mathcal{R}^{\hat\alpha}{}_{\nu}$, we examine the general conditions for gravitational reconnection. 

Since gravitational reconnection requires a breakdown of the ideal Lie transport of the gravitational field-strength two-form, we reconsider the transport of $\mathcal F^{\hat\alpha}$ by the flow generated by $u^\mu$. Repeating the previous steps, but now using the non-ideal Ohm-type condition~\eqref{general_ohm_law}, yields
\begin{equation}
(\mathcal L_u \mathcal F^{\hat\alpha})_{\mu\nu}
=
\nabla_\mu \mathcal R^{\hat\alpha}{}_\nu
- \nabla_\nu \mathcal R^{\hat\alpha}{}_\mu \, .
\label{nonideal_lie}
\end{equation}
Substituting Eq.~\eqref{nonideal_lie} into Eq.~\eqref{eq:flux_transport} gives
\begin{equation}
\frac{d\Phi^{\hat\alpha}}{d\tau}
=
\int_{\mathcal S(\tau)}
\nabla_{[\mu}
\mathcal R^{\hat\alpha}{}_{\nu]}
\,d\Sigma^{\mu\nu} \, .
\label{flux_balance_nonideal}
\end{equation}
Thus, a local breakdown of the ideal Lie transport occurs wherever
\begin{equation}
\nabla_{[\mu}
\mathcal R^{\hat\alpha}{}_{\nu]}
\neq 0 \, ,
\end{equation}
while a net change of gravitational flux through a particular transported surface additionally requires the corresponding surface integral to be nonzero.

However, a local violation of the ideal Lie transport of $\mathcal F^{\hat\alpha}$ with respect to the transport field $u^\mu$ does not by itself imply gravitational reconnection. If there exists a vector field $\zeta^\mu$ such that $\mathcal R^{\hat\alpha}{}_\nu = \mathcal F^{\hat\alpha}{}_{\lambda\nu}\zeta^\lambda$ and the vector field
\begin{equation}
w^\mu \equiv u^\mu-\zeta^\mu
\label{w_def}
\end{equation}
defines a regular transport flow, then the generalized gravitational Ohm-type condition~\eqref{general_ohm_law} becomes
\begin{equation}
w^\lambda \mathcal F^{\hat\alpha}{}_{\lambda\nu} = 0 \, 
\end{equation}
and the ideal transport argument then applies with $u^\mu$ replaced by $w^\mu$. Consequently, the flux of $\mathcal F^{\hat\alpha}$ is conserved along the flow generated by $w^\mu$ and the connectivity of the corresponding gravitational field sheets is preserved. In this case, $\mathcal F^{\hat\alpha}$ slips relative to the original transport field $u^\mu$ but does not reconnect. A necessary condition for gravitational reconnection is therefore 
\begin{equation}
\mathcal{R}^{\hat\alpha}{}_{\nu} \neq  \mathcal{F}^{\hat\alpha}{}_{\lambda\nu}\,\zeta^\lambda 
\label{necessary_reconnection_tensor}
\end{equation}
for every vector field $\zeta^\mu$ such that $w^\mu$ is smooth and nonvanishing.

\begin{figure}
\begin{center}
\includegraphics[width=8.65cm]{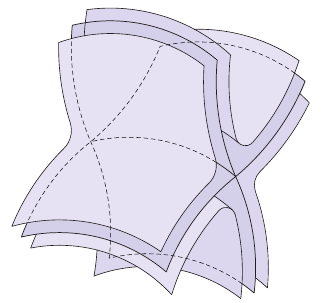}
\end{center}
\vspace{-0.5cm}
\caption{Schematic illustration of neighboring gravitational field sheets associated with the gravitational field-strength two-form $\mathcal F^{\hat\alpha}$. The intersection of separatrix gravitational field sheets defines a separator curve associated with changes in gravitational connectivity.}
\label{fig1}
\end{figure}

To quantify the transfer of gravitational flux, let $\partial\mathcal S$ denote a contour of interest and let $\mathcal S$ be an oriented two-surface spanning it. 
The rate of change of gravitational flux through $\mathcal S$ is
\begin{equation}
\frac{d\Phi^{\hat\alpha}}{d\tau}
=
\int_{\partial\mathcal S} u^\lambda
\mathcal F^{\hat\alpha}{}_{\lambda\nu}
\,dl^\nu \, ,
\label{flux_rate_R}
\end{equation} 
where $dl^\nu$ is the directed line element tangent to $\partial\mathcal S$. 
A nonzero flux transfer rate therefore requires $u^\lambda\mathcal F^{\hat\alpha}{}_{\lambda\nu} \neq 0$ at least somewhere along $\partial\mathcal S$.
Equation~\eqref{flux_rate_R} motivates characterizing gravitational reconnection through line integrals evaluated along the curves where the connectivity of gravitational field sheets changes. The corresponding \emph{separator curves}, $\mathcal C_{\rm sep} = \mathcal M_1\cap\mathcal M_2$, are formed by the intersections of separatrix gravitational field sheets $\mathcal M_1$ and $\mathcal M_2$, which separate regions of distinct gravitational field-sheet connectivity across the non-ideal region (see Fig. \ref{fig1}).

In configurations admitting spatial translational symmetry, the integration surface $\mathcal S$ may be chosen so that only the boundary segment coincident with the separator curve $\mathcal C_{\rm sep}$ contributes to the flux transfer rate. The corresponding gravitational reconnection rate is then given by 
\begin{equation}
\frac{d\Phi_{\rm rec}^{\hat\alpha}}{d\tau}
=
\left|\int_{\mathcal C_{\rm sep}}u^\lambda\mathcal F^{\hat\alpha}{}_{\lambda\nu}\,dl^\nu\right| \, ,
\label{Rg_2D_tensor}
\end{equation}
where $dl^\nu$ is the directed line element tangent to $\mathcal C_{\rm sep}$.

In the absence of spatial translational symmetry, a preferred separator curve is generally not available. Motivated by the definition of the magnetic reconnection rate in plasmas \cite{ComissoJPP_2016}, we consider integral curves $\Upsilon$ of the vector field $u_\lambda\,{}^*\mathcal F^{\hat\alpha\mu\lambda}$ that intersect the non-ideal region and extend into the ideal region. For each such curve, parameterized by a parameter $\lambda$ with directed line element $dl^\nu=(dx^\nu/d\lambda)\,d\lambda$, the accumulated non-ideal contribution along $\Upsilon$ is measured by the line integral of $u^\lambda\mathcal F^{\hat\alpha}{}_{\lambda\nu}$. The gravitational reconnection rate is then defined as
\begin{equation}
\frac{d\Phi_{\rm rec}^{\hat\alpha}}{d\tau}
=
\max_{\Upsilon}
\left|
\int_{\Upsilon}
u^\lambda
\mathcal F^{\hat\alpha}{}_{\lambda\nu}
\,dl^\nu
\right| ,
\label{Rg_3D_tensor}
\end{equation}
where the maximum is taken over the integral curves $\Upsilon$ described above.
Equations~\eqref{Rg_2D_tensor} and~\eqref{Rg_3D_tensor} provide quantitative measures of gravitational reconnection through the transfer of gravitational flux associated with changes in gravitational connectivity.

Gravitational reconnection also affects the exchange of energy and momentum between the Sparling gravitational energy-momentum current and the matter energy-momentum current.
The total energy--momentum current satisfies the conservation law \eqref{cons_current}, whereas the corresponding densitized gravitational and matter currents need not be separately conserved.
Accordingly, Eq.~\eqref{cons_current} can be written as the pair of balance laws
\begin{equation}
\partial_\nu\left(\sqrt{-g}\,t^\nu{}_{\hat\alpha}\right)
= +\mathcal X_{\hat\alpha} \, , \label{eq:balance_sparling}
\end{equation}
\begin{equation}
8\pi\,\partial_\nu\left(\sqrt{-g}\,T^\nu{}_{\hat\alpha}\right)
= -\mathcal X_{\hat\alpha} \, , \label{eq:balance_matter}
\end{equation}
where $\mathcal X_{\hat\alpha}$ denotes the local energy-momentum exchange density between the Sparling gravitational and matter currents.

An explicit expression for the exchange density
$\mathcal X_{\hat\alpha}$ follows by expressing the conservation law $\nabla_\mu T^\mu{}_\nu=0$ in the tetrad frame.
With the Lorentz-covariant derivative convention
\begin{equation}
D_\mu V^\nu{}_{\hat{\alpha}} = \partial_\mu V^\nu{}_{\hat{\alpha}} 
+ \Gamma^\nu{}_{\mu\kappa} V^\kappa{}_{\hat{\alpha}}
+ \omega^{\hat{\beta}}{}_{\hat{\alpha}\mu} V^\nu{}_{\hat{\beta}} \, ,
\end{equation}
the tetrad postulate implies
\begin{equation}
\nabla_\mu \mathcal A^\nu{}_{\hat\alpha} =
-\omega^{\hat\beta}{}_{\hat\alpha\mu}
\mathcal A^\nu{}_{\hat\beta} \, ,
\end{equation}
which, together with the matter conservation law, yields
\begin{equation}
\mathcal X_{\hat\alpha}
=
4\pi\sqrt{-g}\,
\left(\mathcal F_{\hat\alpha}{}^{\hat\beta}{}_{\hat\gamma}
+
\mathcal F_{\hat\gamma}{}^{\hat\beta}{}_{\hat\alpha}
-
\mathcal F^{\hat\beta}{}_{\hat\alpha\hat\gamma}\right)
\mathcal A^{\hat\gamma}{}_\nu T^\nu{}_{\hat\beta} \, .
\label{eq:exchange_density}
\end{equation}
Thus, with the above sign convention, $\mathcal X_{\hat\alpha} < 0$ corresponds to a local gain of the matter current and a local loss of the gravitational current, whereas $\mathcal X_{\hat\alpha} > 0$ corresponds to the opposite transfer.
Since the gravitational field strength $\mathcal F^{\hat\alpha}{}_{\mu\nu}$ enters explicitly in the exchange density $\mathcal X_{\hat\alpha}$, any non-ideal evolution of $\mathcal X_{\hat\alpha}$, including that associated with gravitational reconnection, modifies the local rate of energy-momentum exchange between the gravitational and matter sectors.

Gravitational helicity conservation is another property of the ideal evolution of the gravitational field that is modified by gravitational reconnection. To show this, we define the gravitational helicity current as \cite{AWC}
\begin{equation}
    \mathcal{K}^\mu={^*\mathcal{F}}^{\hat\alpha\mu\nu} \mathcal{A}_{\hat\alpha\nu} \, ,
\label{eq:helicity_current}
\end{equation} 
in analogy with the electromagnetic helicity current, with the electromagnetic field strength and potential replaced by their gravitational counterparts.
Using the homogeneous identity \eqref{homogeneous}, its covariant divergence is
\begin{equation}
\nabla_\mu \mathcal K^\mu
= \frac{1}{2}{^*\mathcal F}^{\hat\alpha\mu\nu}\mathcal F_{\hat\alpha\mu\nu} \, .
\label{eq:helicity_divergence}
\end{equation}
Employing the four-dimensional algebraic identity
\begin{equation}
\mathcal F^{\hat\alpha}{}_{\lambda\mu}\,
{}^*\mathcal F_{\hat\alpha}{}^{\mu\nu}
=
-\frac{1}{4}\delta_\lambda^{\ \nu}\,
\mathcal F^{\hat\alpha}{}_{\rho\sigma}\,
{}^*\mathcal F_{\hat\alpha}{}^{\rho\sigma},
\label{eq:twoform_identity}
\end{equation}
which is satisfied by any antisymmetric two-form, together with the non-ideal gravitational Ohm-type relation \eqref{general_ohm_law}, the covariant divergence becomes
\begin{equation}
\nabla_\mu \mathcal K^\mu
=
2 \, \mathcal R^{\hat\alpha}{}_\mu\, {^*\mathcal F}_{\hat\alpha}{}^{\mu\nu} u_\nu \, .
\label{eq:helicity_balance_nonideal}
\end{equation}
In the ideal limit, \(\mathcal R^{\hat\alpha}{}_\mu=0\), and therefore $\nabla_\mu \mathcal K^\mu=0$. 
In the non-ideal case, however, Eq.~\eqref{eq:helicity_balance_nonideal} shows that gravitational helicity can be locally generated or destroyed through the coupling of the non-ideal term $\mathcal R^{\hat\alpha}{}_\mu$ to ${^*\mathcal F}_{\hat\alpha}{}^{\mu\nu}u_\nu$.

The corresponding gravitational helicity on a spacelike hypersurface $\Sigma(\tau)$, labeled by the proper time $\tau$, is
\begin{equation}
\mathcal H(\tau) = \int_{\Sigma(\tau)} \mathcal K^\mu\, d\Sigma_\mu \, ,
\label{eq:helicity_integral}
\end{equation}
where $d\Sigma_\mu=-n_\mu dV$ is the directed hypersurface element on $\Sigma(\tau)$, with $n^\mu$ the future-directed unit normal and $dV$ the induced volume element. 
Gauge invariance of $\mathcal H$ requires either a closed hypersurface or boundary conditions that ensure the vanishing of the boundary term arising under gauge transformations. 
Applying the covariant divergence theorem to Eq. \eqref{eq:helicity_balance_nonideal} over the spacetime slab bounded by $\Sigma(\tau)$, $\Sigma(\tau+d\tau)$, and the timelike boundary, and assuming vanishing gravitational-helicity flux through the latter, yields 
\begin{equation}
\frac{d\mathcal H}{d\tau}
=
2 \int_{\Sigma(\tau)}
\mathcal R^{\hat\alpha}{}_\mu\,
{}^*\mathcal F_{\hat\alpha}{}^{\mu\nu}
u_\nu\, dV \, .
\label{eq:helicity_integral_balance}
\end{equation}
The ideal limit $\mathcal R^{\hat\alpha}{}_\mu=0$ implies $d\mathcal H/d\tau=0$, so gravitational helicity is conserved. Gravitational reconnection, however, removes this ideal constraint, allowing gravitational helicity to change through the non-ideal coupling $\mathcal R^{\hat\alpha}{}_\mu\,{}^*\mathcal F_{\hat\alpha}{}^{\mu\nu}u_\nu$.

\section{Gravitational Reconnection: Field Decomposition}

To define gravitational analogues of electric and magnetic fields, we perform a $3+1$ decomposition of spacetime \cite{ADM62} with future-directed unit normal $n^\mu$, normalized according to $n_\mu n^\mu=-1$, and the spatial projector
\begin{equation}
\gamma^\mu{}_\nu = g^\mu{}_\nu+n^\mu n_\nu \, .
\end{equation}
The contractions of the gravitational field-strength tensor and its dual with the unit normal $n^\mu$ define ``gravitational electric'' and ``gravitational magnetic'' fields, 
\begin{align}
\mathcal{E}^{\hat{\alpha}\mu} &= n_\lambda\,\mathcal{F}^{\hat{\alpha}\mu\lambda} \, , \label{eq:E_tensor}\\
\mathcal{B}^{\hat{\alpha}\mu} &= n_\lambda\,{^*\mathcal{F}^{\hat{\alpha}\mu\lambda}} \, . \label{eq:B_tensor}
\end{align} 
Accordingly, the ``gravitational electric'' and ``gravitational magnetic'' fields are purely spatial, satisfying $n_\mu \mathcal{E}^{\hat\alpha\mu}=0$ and $n_\mu \mathcal{B}^{\hat\alpha\mu} = 0$. 

With these definitions, the gravitational field-strength tensor and its dual are given by
\begin{align}
{\mathcal{F}^{\hat\alpha\mu\nu}}
&=
n^\mu \mathcal{E}^{\hat\alpha\nu} 
-
 n^\nu \mathcal{E}^{\hat\alpha\mu}
-
\epsilon^{\mu\nu\rho\lambda}
{\mathcal{B}^{\hat\alpha}}_{\rho}
n_\lambda \, ,
\label{expanF1}
\\
{^*\mathcal{F}}^{\hat\alpha\mu\nu}
&=
-\mathcal{B}^{\hat\alpha\mu} n^\nu
+
\mathcal{B}^{\hat\alpha\nu} n^\mu
-
\epsilon^{\mu\nu\rho\lambda}
{\mathcal{E}^{\hat\alpha}}_{\rho}
n_\lambda \, ,
\label{expanF2}
\end{align}
where $\epsilon^{\mu\nu\rho\lambda}$ denotes the Levi--Civita tensor. The transport four-velocity is decomposed with respect to the same foliation as
\begin{equation}
u^\mu =
\Gamma\left(n^\mu+v^\mu\right) \, ,
\qquad
n_\mu v^\mu=0 \, ,
\label{velocitydefinition}
\end{equation}
where $v^\mu$ is the hypersurface-tangent component of the four-velocity and
\begin{equation}
\Gamma = -\,n_\mu u^\mu
\end{equation}
is the Lorentz factor measured by the Eulerian observer associated with the foliation.

Using the decomposition \eqref{expanF1} and Eq.~\eqref{velocitydefinition}, the non-ideal gravitational Ohm-type relation \eqref{general_ohm_law} becomes
\begin{equation}
(u^\mu {\mathcal E}^{\hat\alpha}{}_\mu)n_\nu
+ \Gamma {\mathcal E}^{\hat\alpha}{}_\nu
+ u^\mu \epsilon_{\mu\nu\rho\lambda}
\mathcal B^{\hat\alpha\rho} n^\lambda
= -{\mathcal R}^{\hat\alpha}{}_\nu \, .
\end{equation}
Contracting this equation with $n^\nu$ and using $n^\nu{\mathcal E}^{\hat\alpha}{}_\nu=0$ and $n_\nu n^\nu=-1$ gives
\begin{equation}
u^\mu{\mathcal E}^{\hat\alpha}{}_\mu
=
 n^\mu{\mathcal R}^{\hat\alpha}{}_\mu \, .
\end{equation}
Since $u^\nu{\mathcal R}^{\hat\alpha}{}_\nu=0$, the velocity
decomposition \eqref{velocitydefinition} implies $n^\nu{\mathcal R}^{\hat\alpha}{}_\nu = - v^\nu{\mathcal R}^{\hat\alpha}{}_\nu$, from which it follows that $u^\mu{\mathcal E}^{\hat\alpha}{}_\mu = -v^\nu{\mathcal R}^{\hat\alpha}{}_\nu$.
Thus, unlike in the ideal case discussed in Ref.~\cite{AWC},
$u^\mu{\mathcal E}^{\hat\alpha}{}_\mu$ need not vanish. 
Projecting onto the hypersurface orthogonal to $n^\mu$, the non-ideal gravitational Ohm-type relation becomes
\begin{eqnarray}
{\mathcal{E}^{\hat\alpha}}_\mu&=&\frac{1}{\Gamma}\epsilon_{\mu\nu\rho\lambda}u^\nu \mathcal{B}^{\hat\alpha \rho}n^\lambda - \frac{1}{\Gamma} {\gamma_\mu}^{\nu} {\mathcal{R}^{\hat\alpha}}_{\nu} \, . \label{nonidealC}
\end{eqnarray} 
This relation contains a non-ideal contribution that cannot be written as a pure cross product with $\mathcal B^{\hat\alpha\mu}$. Contracting Eq.~\eqref{nonidealC} with
\({\mathcal B}_{\hat\alpha}{}^\mu\) yields
\begin{equation}
{\mathcal{B}_{\hat\alpha}}^\mu {\mathcal{E}^{\hat\alpha}}_\mu = -\frac{1}{\Gamma} {\mathcal{B}_{\hat\alpha}}^\mu  {\mathcal{R}^{\hat\alpha}}_{\mu}\, ,
\label{BE_nonideal}
\end{equation}
where the right-hand side generally does not vanish, in contrast to the ideal case \cite{AWC}.

Introducing coordinates $(t,x^i)$ adapted to the foliation, with the hypersurfaces of the foliation given by $t=\mathrm{const}$, the unit normal has components
\begin{align}
n_\mu &= (-\alpha,0_i) \, , \\
n^\mu &= (1/\alpha,-\beta^i/\alpha) \, ,
\end{align} 
where $\alpha$ is the lapse function and $\beta^i$ is the shift vector. 
Expressing Eq.~\eqref{nonidealC} in coordinates adapted to the foliation gives
\begin{equation}
\alpha\,{\boldsymbol{\mathcal E}}^{\hat\alpha}
+
(\boldsymbol v+\boldsymbol\beta)
\times
{\boldsymbol{\mathcal B}}^{\hat\alpha}
=
-\frac{\alpha}{\Gamma}
{\boldsymbol{\mathcal R}}^{\hat\alpha} \, ,
\label{nonideal_Ohm_vectorial}
\end{equation}
where $\boldsymbol v$ is the three-vector whose coordinate components are $(\boldsymbol v)^i=\alpha u^i/\Gamma$, with $\Gamma=\alpha u^0$, and ${\boldsymbol{\mathcal R}}^{\hat\alpha}$ is the three-vector associated with the hypersurface projection $\gamma_\mu{}^\nu{\mathcal R}^{\hat\alpha}{}_\nu$. Here and in the following, the vector product is defined using the three-dimensional alternating symbol.

Expressing Eq.~\eqref{homogeneous} in coordinates adapted to the foliation similarly gives
\begin{equation}
\frac{\partial}{\partial t}\left(\sqrt{\gamma}\, {\boldsymbol{\mathcal{B}}}^{\hat\alpha} \right)+\nabla\times\left(\alpha{\boldsymbol{\mathcal{E}}}^{\hat\alpha} + \boldsymbol{\beta}\times{\boldsymbol{\mathcal{B}}}^{\hat\alpha}\right)=0 \, ,
\label{homogeneous_vector}
\end{equation}
where $\gamma=\det(\gamma_{ij})$, so that $\sqrt{\gamma}=\sqrt{-g}/\alpha$, and $\nabla\times$ denotes the coordinate curl, $(\nabla\times\boldsymbol X)^i \equiv [ijk]\,\partial_jX_k$, with $[ijk]$ the three-dimensional alternating symbol.
Combining Eqs.~\eqref{nonideal_Ohm_vectorial} and \eqref{homogeneous_vector}, we obtain
\begin{equation}
\frac{\partial}{\partial t}
\left(
\sqrt{\gamma}\,
{\boldsymbol{\mathcal B}}^{\hat\alpha}
\right)
-
\nabla\times
\left(
{\boldsymbol v}
\times
{\boldsymbol{\mathcal B}}^{\hat\alpha}
\right)
=
\nabla\times \left( \frac{\alpha}{\Gamma}
{\boldsymbol{\mathcal R}}^{\hat\alpha} \right)  \, .
\label{nonideal_induction_vector}
\end{equation}
The second term on the left-hand side represents the ideal advection of the ``gravitational magnetic'' field, whereas the term on the right-hand side represents the non-ideal contribution.

This decomposition motivates the definition of a local Reynolds-like number for the ``gravitational magnetic'' field as the ratio of the advection and non-ideal terms appearing in Eq.~\eqref{nonideal_induction_vector}: 
\begin{equation}
{\Xi}^{\hat\alpha}_{\mathcal R} \equiv \frac{ \left\| \nabla\times\left(\boldsymbol{v}\times{\boldsymbol{\mathcal B}}^{\hat\alpha}\right) \right\| }{ \left\| \nabla \times \left( \dfrac{\alpha}{\Gamma}
{\boldsymbol{\mathcal R}}^{\hat\alpha} \right) \right\| } \, .
\label{Re_g_def_vector}
\end{equation}
Here $\|\cdot\|$ denotes the norm induced by the spatial metric $\gamma_{ij}$.
Values ${\Xi}^{\hat\alpha}_{\mathcal R} \sim\mathcal O(1)$ indicate that the non-ideal contribution is locally comparable to the ideal transport contribution and therefore cannot be neglected in the evolution of the ``gravitational magnetic'' field. 
The presence of a significant non-ideal contribution alone is not sufficient for gravitational reconnection, which additionally requires that the non-ideal term cannot be eliminated by a redefinition of the spatial transport field, i.e. ${\boldsymbol{\mathcal R}}^{\hat\alpha} \neq \boldsymbol{\zeta}\times{\boldsymbol{\mathcal{B}}}^{\hat\alpha}$ (see condition \eqref{necessary_reconnection_tensor}). 

The field decomposition discussed above also allows for a convenient representation of the gravitational flux and reconnection measures introduced in Sec.~\ref{sec:grav_reconn}. 
In particular, the gravitational flux \eqref{flux_tensor_def} can be expressed in terms of the ``gravitational magnetic'' field as
\begin{equation}
\Phi^{\hat\alpha}
=
\int_{\mathcal S}
ds\,\sqrt{\gamma}\,
{\mathcal B}^{\hat\alpha\rho}
s_\rho \, ,
\label{MagnefluxesVector}
\end{equation}
where the directed surface element on the hypersurfaces of the foliation is related to the antisymmetric surface element in Eq.~\eqref{flux_tensor_def} by $\sqrt{\gamma}\, {{s}}_{\rho}\,ds=\epsilon_{\mu\nu\rho\lambda}n^\lambda d \Sigma^{\mu\nu}$.
Therefore, the flux of the ``gravitational magnetic'' field through the two-surface $\mathcal S$ comoving with $\boldsymbol v$ evolves according to
\begin{equation}
\frac{d\Phi^{\hat\alpha}}{dt} = \int_{\mathcal S} ds \, \left(\nabla\times \dfrac{\alpha}{\Gamma}
{\boldsymbol{\mathcal R}}^{\hat\alpha} \right) \cdot{\boldsymbol{s}}\, .
\label{flux_evolution_nonideal}
\end{equation}

The reconnection measures given by Eqs.~\eqref{Rg_2D_tensor} and \eqref{Rg_3D_tensor} can likewise be expressed in vector form.
In configurations admitting a spatial translational symmetry, the integration contour $\partial\mathcal S$ may be chosen so that only segments parallel to the translationally invariant direction $\hat{\boldsymbol{\zeta}}$ contribute. 
Equation~\eqref{Rg_2D_tensor} then reduces to
\begin{equation}
\frac{d\Phi_{\rm rec}^{\hat\alpha}}{dt}
=
\left| \int_{\mathcal C_{\rm sep}}
dl_{\hat\zeta}\,
\left(\alpha{\boldsymbol{\mathcal E}}^{\hat\alpha}
+\boldsymbol{\beta}\times
{\boldsymbol{\mathcal B}}^{\hat\alpha}\right)
\cdot \hat{\boldsymbol{\zeta}} \right| \, .
\label{Rg_2D_vector}
\end{equation}

In the absence of spatial translational symmetry, we consider integral curves $\Upsilon$ of the ``gravitational magnetic'' field ${\boldsymbol{\mathcal B}}^{\hat\alpha}$ that thread the non-ideal region, where ${\boldsymbol{\mathcal E}}^{\hat\alpha}\!\cdot {\boldsymbol{\mathcal B}}^{\hat\alpha}\neq 0$, and extend into ideal regions, where ${\boldsymbol{\mathcal E}}^{\hat\alpha}\!\cdot {\boldsymbol{\mathcal B}}^{\hat\alpha}=0$. 
Parameterizing each curve by arclength $l$, the corresponding directed line element is $d\boldsymbol{l} = ({{\boldsymbol{\mathcal B}}^{\hat\alpha}}/{\lvert{\boldsymbol{\mathcal B}}^{\hat\alpha}\rvert})dl$.
Equation~\eqref{Rg_3D_tensor} then becomes
\begin{equation}
\frac{d\Phi_{\rm rec}^{\hat\alpha}}{dt}
=
\max_{\Upsilon}
\left|
\int_{\Upsilon} dl\,\alpha \,
\frac{{\boldsymbol{\mathcal E}}^{\hat\alpha} \cdot {\boldsymbol{\mathcal B}}^{\hat\alpha}}{\lvert{\boldsymbol{\mathcal B}}^{\hat\alpha}\rvert} \right| \, .
\label{Rg_3D_vector}
\end{equation}
Equations~\eqref{Rg_2D_vector} and \eqref{Rg_3D_vector} show that gravitational reconnection can be quantified entirely in terms of the ``gravitational electric'' and ``gravitational magnetic'' fields.

The gravitational helicity and its evolution also admit a representation in terms of the ``gravitational magnetic'' field. Using the decompositions \eqref{expanF1} and \eqref{expanF2}, the local gravitational helicity balance law takes the form
\begin{equation}
\nabla_\mu \mathcal K^\mu
= -2{\mathcal{B}_{\hat\alpha}}^\mu {\mathcal{E}^{\hat\alpha}}_\mu = \frac{2}{\Gamma} {\boldsymbol{\mathcal R}}^{\hat\alpha}\cdot {\boldsymbol{\mathcal B}}_{\hat\alpha}\, ,
\label{eq:helicity_balance_nonideal6}
\end{equation}
where the second equality follows from Eq.~\eqref{BE_nonideal}. The gravitational helicity \eqref{eq:helicity_integral} then becomes
\begin{equation}
    \mathcal{H} = -\int_{\Sigma_t} d^3x\,\sqrt{\gamma}\, n^\mu\mathcal K_\mu = \int_{\Sigma_t} d^{3}x \, \sqrt{\gamma} \, {\boldsymbol{\mathcal{A}}}_{\hat\alpha} \cdot {\boldsymbol{\mathcal{B}}}^{\hat\alpha} \, ,
\end{equation}
where we used $d\Sigma^\mu=-n^\mu\sqrt{\gamma} \, d^3x$, so that $\alpha n_\mu d\Sigma^\mu\,dt = \sqrt{-g}\,dt\,d^3x$ is the spacetime volume element. The corresponding evolution equation for the gravitational helicity then takes the form
\begin{equation}
\frac{d\mathcal H}{d t}
= - 2 \int_{\Sigma_t}
d^3 x \,\alpha \sqrt{\gamma} \,  {\boldsymbol{\mathcal B}}_{\hat\alpha} \cdot {\boldsymbol{\mathcal E}}^{\hat\alpha} = 2 \int_{\Sigma_t}
d^3 x \,\alpha \sqrt{\gamma} \, \frac{ {\boldsymbol{\mathcal R}}^{\hat\alpha}\cdot {\boldsymbol{\mathcal B}}_{\hat\alpha}}{\Gamma}
\label{eq:helicity_integral_balance8}
\end{equation}
up to the helicity flux through the boundary of the integration domain.
Thus, gravitational helicity is generated or annihilated by the component of the non-ideal field parallel to the ``gravitational magnetic'' field.

\section{Conclusions}

In an ideal evolution of the gravitational field, characterized by the ideal gravitational Ohm-type condition $u^\lambda\mathcal F^{\hat\alpha}{}_{\lambda\nu}=0$, the gravitational flux $\Phi^{\hat\alpha}$ associated with the gravitational field-strength tensor $\mathcal F^{\hat\alpha}{}_{\mu\nu}$ is conserved and the connectivity of the corresponding gravitational field sheets is preserved. 
These properties impose topological constraints on the admissible spacetime dynamics. Gravitational reconnection becomes possible only when these constraints break down.

Departures from ideal gravitational evolution arise either from singularities in the transport field $u^\mu$ or from a violation of the ideal gravitational Ohm-type condition, $u^\lambda \mathcal F^{\hat\alpha}{}_{\lambda\nu}\neq0$, which may reflect the absence of an ideal transport congruence, effective coarse graining, or modifications of general relativity. 
We have shown that violating the ideal gravitational Ohm-type condition is not by itself sufficient to change the connectivity of gravitational field sheets. If the non-ideal term can be absorbed into a redefinition of the transport field, gravitational flux freezing is restored with respect to the redefined transport field and the frozen-in connectivity is preserved. Changes in gravitational connectivity therefore occur only when no such redefinition is possible.

To quantify gravitational reconnection, we have introduced measures of the gravitational reconnection rate, $d\Phi_{\rm rec}^{\hat\alpha}/d\tau$, which characterize the rate at which gravitational flux undergoes reconnection in both symmetry-reduced and general configurations.
These measures provide quantitative diagnostics of gravitational reconnection and a means of comparing reconnection processes across different configurations.

Gravitational reconnection has consequences that extend beyond changes in gravitational field-sheet connectivity. We have shown that it alters the exchange of energy and momentum between the Sparling gravitational and matter energy-momentum currents through the exchange density $\mathcal X_{\hat\alpha}$ and gives rise to changes in the gravitational helicity $\mathcal H$.
Gravitational reconnection therefore constitutes a mechanism that simultaneously reshapes gravitational connectivity, alters the local exchange of energy and momentum between the gravitational and matter currents, and modifies gravitational helicity.

Upon decomposing the gravitational field-strength tensor $\mathcal F^{\hat\alpha}{}_{\mu\nu}$ into ``gravitational electric'' and ``gravitational magnetic'' fields, these results admit a formulation closely analogous to magnetohydrodynamics.
The gravitational field sheets correspond to ``gravitational magnetic'' field sheets, whose intersections with the spacelike hypersurfaces of the foliation define ``gravitational magnetic'' field lines. 
The corresponding evolution equations take a magnetohydrodynamic-like form. 
From these equations, we define a local Reynolds-like number, $\Xi^{\hat\alpha}_{\mathcal R}$, which quantifies the relative importance of the advective and non-ideal terms. 
The same decomposition also enables a magnetohydrodynamic-like interpretation of the gravitational helicity evolution law.

The framework developed here provides a means of characterizing changes in gravitational connectivity and determining when the topological constraints associated with ideal gravitational evolution break down.
The existence of these topological constraints, together with the possibility of gravitational reconnection, suggests that gravitational connectivity plays an important role in nonlinear spacetime dynamics. 
The present results form a basis for investigating changes in gravitational connectivity, energy-momentum exchange, and gravitational helicity in nonlinear regimes of general relativity.

$\,$

\begin{acknowledgments}
We thank Filippo Camilloni, Luciano Combi, and Peter Rau for insightful comments on the manuscript. 
L.C. acknowledges support from NSF grant PHY-2308944 and NASA ATP award 80NSSC24K1230.
F.A.A. acknowledges support from FONDECYT grant No. 1230094. 
\end{acknowledgments}

\end{document}